\documentstyle{mn}

\def\text{\rm}
\def\kms{km~s$^{-1}$}

\def\cm2{cm$^{-2}$}

\def\lya{{\rm Ly}$\alpha$}

\newcommand{\lsim}{\ \raise -2.truept\hbox{\rlap{\hbox{$\sim$}}\raise5.truept
        \hbox{$<$}\ }}
\newcommand{\gsim}{\ \raise -2.truept\hbox{\rlap{\hbox{$\sim$}}\raise5.truept
        \hbox{$>$}\ }}

\begin{document}

\title{The power spectrum of the Lyman-$\alpha $ clouds }
\author[Luca Amendola \& Sandra Savaglio  ]{Luca Amendola$^1$ \&
 Sandra Savaglio$^{2,3}$\\ 
 $^1$Osservatorio Astronomico di Roma, 
Viale del Parco Mellini 84, 00136 Roma. Italy,
amendola@oarhp1.rm.astro.it\\
$^2$ European Southern Observatory, Karl-Schwarzschildstr. 2,
        D--85748 Garching bei M\"unchen, Germany\\
$^3$ Space Telescope Science Institute, 3700 San Martin Drive,
        Baltimore, MD21218, USA}
\date{Accepted: April 1999 }
\maketitle

\begin{abstract}
We investigate the clustering properties of 13 QSO lines of sight in
flat space, with average redshifts from $z\approx 2$ to 4. We estimate
the 1--D power spectrum and the integral density of neighbours, and
discuss their variation with respect to redshift and column
density. We compare the results with standard CDM models, and estimate
the power spectrum of Lyman-$\alpha $ clustering as a function both of
redshift and column density.  We find that $a$) there is no
significant periodicity or characteristic scale; $b$) the clustering
depends both on column density and redshift; $c$) the clustering
increases linearly only if at the same time the HI column density
decreases strongly with redshift. The results remain qualitatively the
same assuming an open cosmological model.

\end{abstract}

\input epsf.tex

\section{Introduction}
 
As new and deeper galaxy redshift surveys are being completed, a
definite picture of the local galaxy distribution is slowly
emerging. This distribution is characterized by large voids surrounded
by sheets of clustered matter (see e.g. El-Ad et al. 1996a, 1997), and
can be well quantified by statistical descriptors like the power
spectrum (see e.g. Park et al. 1994, Tadros \& Efstathiou 1995, Tadros
\& Efstathiou 1996, Lin et al.  1996). In a few years, surveys like
the Sloan Digital Sky Survey (Loveday 1996) and the Two Degree Field
Redshift Survey (Colless 1998, Maddox 1998) will extend our knowledge
of the clustering of the luminous matter by at least a factor of ten
both in depth and coverage.
 
However, mapping the luminous galaxies is clearly not enough to fully
understand the history and geography of our Universe. First, even the
deepest surveys so far planned do not reach beyond $z\approx
1$. Secondly, we already know that a prominent component of the
matter, perhaps a dominant one, is not bright enough to be included in
the present surveys. Ranging from truly dark matter, to dwarf
galaxies, to very low surface brightness galaxies, to Lyman-$\alpha $
clouds, the systems which can easily escape detection are many and
varied.  Their importance in solving crucial questions, such as
whether the voids traced by bright galaxies are really empty of matter
(Szomoru et al. 1994, Stocke et al. 1995), which are the paths of
galaxy formation, what is the role of galaxy interactions, cannot be
underestimated.
 
In this respect, the large number of absorption lines seen in quasar
spectra, the so called {\rm Ly}$\alpha $~forest, is an important
tracer of the intervening matter distribution along the QSO lines of
sight at any redshift in the observable Universe ($0\lsim z \lsim
5$). Even if the {\rm Ly}$\alpha $~forest is a one-dimensional
distribution, the fact that a line of sight can contain at high
redshift ($z\ \raise -2.truept
\hbox{\rlap{\hbox{$\sim$}}\raise5.truept \hbox{$>$}\ }2.5$) up to a
few hundreds of absorption lines makes the statistics
significant. While galaxies are associated with emitting objects, the
{\rm Ly}$\alpha $~clouds represent the gas detected through the
absorption of HI and are not necessarily associated with stars. In
this sense they are matter tracers different from any other.
 
What is the fraction of baryonic matter in the IGM is still matter of
controversy.  Hydrodynamical simulations of the {\rm Ly}$\alpha
$~forest (Rauch \& Haehnelt 1995, Miralda--Escud\`{e} et al. 1996,
Rauch et al. 1997, Weinberg et al. 1997, Bi \& Davidsen 1997) indicate
that this fraction is large for redshift $z\ \raise -2.truept
\hbox{\rlap{\hbox{$\sim$}}\raise5.truept \hbox{$>$}\ }2$
($\Omega_{b,{\rm IGM } }h^2\ \raise -2.truept
\hbox{\rlap{\hbox{$\sim$}}\raise5.truept \hbox{$>$}\ }0.01-0.02$) and
collapsed objects in the young Universe represent probably a small
correction. However observations of the {\rm Ly}$\alpha $~forest with
column density in the range $12.8\ \raise -2.truept
\hbox{\rlap{\hbox{$\sim$}}\raise5.truept \hbox{$<$}\ }\log N_{{\rm
HI}}\leq 16.0$ by Kim et al. 1997 using Keck/HIRES QSO spectra give
$\Omega_{b,{\rm IGM}}h^2\ \raise -2.truept
\hbox{\rlap{\hbox{$\sim$}}\raise5.truept \hbox{$<$}\ }0.01$ in $2.1 <
z < 3.5$. The discrepancy is probably due to the fact that in the
simulations the (photo and collisional) ionisation of the IGM is
assumed to be higher than that assumed by Kim et al., who neglected
collisional ionisation because they derived an upper limit to the
temperature of the gas $T<10^5$ K from the measured Doppler parameter
of the {\rm Ly}$\alpha$ lines.
 
The study of signatures in the distribution of {\rm Ly}$\alpha $~lines
has been performed mainly using the traditional tool for the galaxy
distribution: the two point correlation function (TPCF). While it has
been clear for many years (Sargent et al. 1980) that low resolution
spectroscopy was unable to give a definitive answer to the clustering
of the {\rm Ly}$\alpha $~forest for scales of $\Delta v<300$
km~s$^{-1}$ (no signal was found for larger scales), the advent of
high resolution spectroscopy in the last few years has started a more
controversial discussion on the presence of clustering at small
scales, and on the evolution of the same with redshift. A weak signal
has been found on small scales ($50<\Delta v<300$ km~s$^{-1}$, Webb
1987, Rauch et al. 1992, Chernomordik 1995, Cristiani et
al. 1995). More recently, a detection of redshift evolution of the
TPCF, being stronger at lower redshift, in a large sample of high
column density {\rm Ly}$\alpha $~lines ($\log N_{{\rm HI}}>13.8$)
(Cristiani et al. 1997, Savaglio et al. 1999) has been only marginally
confirmed by a similar study by Kim et al. (1997), keeping open the
question whether the {\rm Ly}$\alpha $~forest does cluster and at
which scales. This controversy leads to the natural consequence that
it is necessary to find other tools of investigation for the same
problem since the TPCF is neither the only statistical technique nor
necessarily the most appropriate. Pando \& Fang (1996) have shown that
the scaling properties of the {\rm Ly}$\alpha $~forest can be studied
using the wavelet decomposition analysis. The results are that
clustering is present up to large scales (20 $h^{-1}$ Mpc) and that
the 1--D power spectrum is significantly different from a Poissonian
distribution. The box--counting technique (Carbone \& Savaglio 1996,
Savaglio \& Carbone 1997) has the advantage of avoiding boundary
effects, and has given a positive signal evolving with redshift (up to
about 3 $h^{-1}$ Mpc at $z\sim 3.8$ and up to about 20 $h^{-1}$ Mpc at
$z\sim 2$).  The Ly$\alpha$ forest has been studied also by Hui et
al. (1997), by deriving the expected column density distribution given
a theoretical power spectrum.
 
We will investigate the distribution of the Lyman-$\alpha $ clouds by
two different, but related, descriptors: the 1-dim power spectrum, and
the density of neighbours, or conditional density. The former is
particularly suited to detecting preferential scales in the
distributions, since it decomposes the system in plane waves. Here
"preferential scale" means simply a wavelength whose Fourier
coefficient is significatively different than expected assuming a
Gaussian distribution of the Fourier coefficients.  A preferential
scale of clustering has sometimes been claimed in the galaxy
distribution (Broadhust et al. 1991; Einasto et al. 1997), so that it
is interesting to see if something similar holds for the
Lyman-$\alpha$ clouds.  The 1-dim power spectrum is however
intrinsically a very noisy quantity, since the Fourier coefficients
are statistically independent quantities in the limit of a Gaussian
field. The main advantage is that, as a consequence of the statistical
independence, their statistics is straightforward. On the other hand,
the conditional density is an integral quantity, because it counts the
average density of clouds within a distance $R$ of another cloud; its
signal is stable, but its statistics can only be approximated via
MonteCarlo simulations. The conditional density has a reduced scatter
with respect to the TPCF, and is therefore easier to compare to
theoretical models, although the information on specific scales is
spread over a large interval.
 
Once the analysis of the dataset is completed, we can compare it to
theoretical models of the density fluctuation field. The primary goal
is to derive the Lyman-$\alpha $ power spectrum in the linear regime
as a function {\it both of redshift and column density}. We will
therefore derive a biasing function $b_{{\rm {Ly}\alpha } }^2(z,
N_{HI})$, defined as the ratio of the cloud spectrum to a theoretical
CDM spectrum at scales large enough to be in the linear regime of
gravitational clustering. Since we will consider clouds at $z\ge 2$,
we can assume safely that comoving scales larger than 3 Mpc/h are
linear, i.e. the variance in spheres of 3 Mpc/h at $z\ge 2$ is less
than unity. Although we will report data even at lower scales, all the
relevant numerical results are obtained using only scales larger than
3 Mpc/h. For simplicity, we normalize the CDM spectrum to the value
that matches the present day cluster abundance, i.e. $\sigma
_8(z=0)\approx 0.6 \Omega_0^{-0.6}$ (White, Efstathiou \& Frenk 1993);
a different choice of $\sigma _8$, for instance normalizing to the
microwave background fluctuations or to the present galaxy clustering,
simply rescale $b_{{\rm {Ly}\alpha } }^2(z,N_{HI})$ by a constant
factor. Since at the high redshifts of our data the space-time
geometry has an important effect, we test the dependence of our
results considering in the following two values of the cosmological
density: the inflationary value $\Omega_0=1$ and the observationally
preferred $\Omega_0=0.4$.  The final product is the cloud power
spectrum approximated as
\begin{equation}
P_{{\rm Ly}\alpha }(k,z)=b_{{\rm Ly}\alpha }^2(z,N_{HI})P_{cdm}(k,z)
\end{equation}
where $P_{cdm}(k,z)$ is a CDM spectrum normalized to $\sigma _8=0.6
\Omega_0^{-0.6}$, evolving with $z$ as $D_+(\Omega_0,z)^{2}$, where
$D_+(\Omega_0,z)$ is the fluctuation growth function in the linear
regime. We adopt the approximation (Padmanabhan 1993)
\begin{equation}
D_+=\frac{1+\frac{3}{2}\Omega_0}{1+\frac{3}{2}\Omega_0+\frac{5}{2}\Omega_0 z}
\end{equation}
which is valid for $\Omega_0>0.1$ and $\Lambda=0$ (the effect of a finite
$\Lambda$ is actually negligible on $D_+$). For $\Omega_0=1$
this gives $(1+z)^{-1}$ as expected.  The biasing function $b_{{\rm
Ly} \alpha }^2$ expresses therefore the evolution of the clustering of
the clouds both versus redshift and column density. One can
disentangle the evolutionary path in $z$ and $ N_{HI}$  only by means
of further assumptions. Although this task is left to future work, we
obtain the preliminary result that the Ly$ \alpha $ clustering evolves
linearly only if at the same time the average column density decreases
strongly with time.

Our method is based on the parametric estimation of the power spectrum
of fitted absorption lines. A completely different method has been
proposed in a recent work by Croft et al. (1998). They can successfully
recover both the shape and the amplitude of the linear power spectrum
of initial mass fluctuations (assumed to be Gaussian) from the
one--dimensional power spectrum derived from QSO Lyman--$\alpha$
spectra, artificially created by realistic hydrodynamic cosmological
simulations. It is premature to compare the results of the two
approaches, also because Croft et al. apply the method to only one
observed QSO line of sight; the results on the spectrum of Q1244+231
at $z\approx 3$ are consistent with a standard CDM model with
normalization $\sigma_8=0.5$. This conclusion applies also to our
findings, although we find a very marked dependence on redshift and on
column density.

\begin{table*}
\caption[t1]{The QSO absorption line sample}
\label{t1}
\begin{center}
\begin{tabular}{llccccccccc}
\hline\hline
&  &  &  &  &  &  &  &  &  &  \\[-5pt] 
\# & QSO & $z_{em}$ & FWHM & $<z>$ & $\Delta z$ & No. of & $n_{\geq 13.3}$ & 
$n_{\geq 13.5}$ & $n_{\geq 13.8}$ & $n_{\geq 14}$ \\ 
&  &  & (\kms) &  &  & lines &  &  &  &  \\[2pt] \hline
&  &  &  &  &  &  &  &  &  &  \\[-8pt] 
1 & $0000-26^1$ & 4.124 & 6.6 & 3.73 & $3.43-4.04$ & 366 & 248 &206  &142  & 105 \\ 
$1b$ & $0000-26^2$ & 4.124 & 13 & 3.82 & $3.60-4.04$ & 262 &198  &165  &116  &89  \\ 
2 & $1208+10^3$ & 3.82 & 9 & 3.65 & $3.55-3.75$ & 118 &77  &65  &44  &31  \\ 
3 & $0055-26^4$ & 3.67 & 13 & 3.28 & $2.96-3.61$ & 285 & 212 &171  &109  &67  \\ 
4 & $2355+01^3$ & 3.39 & 9 & 3.22 & $3.12-3.34$ & 97 & 68 & 64 & 48 & 33 \\ 
5 & $0014+81^5$ & 3.41 & 8 & 2.95 & $2.71-3.19$ & 262 &122  &96  &66  &48  \\ 
6 & $2126-15^6$ & 3.27 & 11 & 2.90 & $2.60-3.21$ & 154 & 122 & 101 &71  & 55 \\ 
7 & $0302-00^5$ & 3.29 & 8 & 2.87 & $2.63-3.11$ & 266 & 118 & 96 & 62 & 42  \\ 
8 & $0956+12^5$ & 3.30 & 8 & 2.86 & $2.63-3.09$ & 256 & 110  & 81  &48  &40  \\ 
9 & $0636+68^5$ & 3.17 & 8 & 2.79 & $2.55-3.03$ & 313 & 142 & 112 & 68 & 47 \\ 
10 & $1700+64^7$ & 2.72 & 15 & 2.41 & $2.15-2.68$ & 85 & 73  &65  &43  &30  \\ 
11 & $1225+31^8$ & 2.219 & 18 & 1.94 & $1.70-2.18$ & 130 &79  &60  &29  & 16 \\ 
12 & $1101-26^9$ & 2.15 & 9 & 1.97 & $1.85-2.10$ & 65 & 38 &31  &21  &12  \\ 
13 & $1331+17^{10}$ & 2.08 & 18 & 1.87 & $1.68-2.05$ & 57 &42  &31  &19  & 10 \\%
[2pt] \hline
&  &  &  &  &  &  &  &  &  &  \\[-8pt] 
& total &  &  &  & $1.85-4.12$ & 2454 &  &  &  &  \\[2pt] \hline
&  &  &  &  &  &  &  &  &  & 
\end{tabular}
\end{center}
\par
{\footnotesize {$^1$ Lu et al. (1996), $^{2}$ Savaglio et al. (1997), $^3$
Cristiani et al. (1997), $^4$ Cristiani et al. (1995), $^5$ Hu et al. (1995), $
^6$ D'Odorico et al. (1997), $^7$ Rodr\'\i guez--Pascual et al. (1995), $^8$
Khare et al. (1996), $^9$ Carswell et al. (1991), $^{10}$ Kulkarni et al.
(1996)} }
\end{table*}

\section{The sample}

The sample contains 13 absorption line lists of QSO lines of sight
collected from the literature, for a total of 2450 {\rm Ly}$\alpha
$~absorption lines in the redshift range $1.68<z<4.04$ (in the
following we refer to {\it lines} meaning always Ly$\alpha$ absorption
lines).  These line lists have been obtained by means of high
resolution spectroscopy ($6.6\leq $ FWHM $\leq 18$ km~s$^{-1}$) and
have no observational gaps in the {\rm Ly}$\alpha $~forest. The region
of the proximity effect (closer than about 8 Mpc from the source) has
been excluded for completeness. We also excluded from the sample those
systems which are known to be metal systems. In Table \ref{t1} we give
details and references of the sample used. For one object (QSO 1 and
1$^b$ in Table \ref{t1}), we compared two line lists given by two
different groups (Lu et al. 1997, Savaglio et al 1997) in order to
understand whether any significant difference between the two
distributions is connected to different absorption line analysis
approaches; in the analysis we always use the sample 1.

To investigate the dependence of the clustering on the column density
and on the redshift, we group the 13 lines of sight in various
subsets. First, we consider separately the six lines of sight with
higher average redshift (average redshift $ <z>\simeq 3.28$ ; numbers
$1-6$ in Table \ref{t1}), and the seven systems with lower average
redshifts ( $<z>\simeq 2.39;$ numbers $7-13$ in Table \ref{t1}).  We
will refer to these two sets as the higher and the lower redshift
sets, respectively.  Then, we group the 13 QSO lists in six sets of
four lines of sight each (the sets are partially overlapping), so as
to obtain average redshifts ranging from 2.05 to 3.47, as detailed in
Table \ref{t2}. For instance, the number-of-neighbours statistics for
the set $a$ of Table \ref{t2} is obtained averaging over the QSO lines
of sight 0000-26,1208-26,0055-26 and 2355+01.  Finally, we consider
for each of the sets the Lyman-$\alpha$ lines with column density
above a certain threshold, $\log N_{HI}\geq 13.3,13.5,13.8,14$.  In
all, we obtain 30 different, although partially overlapping, subsets
of clouds with various values of cut-off in the column density, and
various average redshifts. This grid of sets will be exploited to get
the full, bidimensional variation of clustering with column density
and redshift.

\section{The method}

\begin{table}
\caption[t1]{Data subsets }
\label{t2}
\begin{center}
\begin{tabular}{ccc}
\hline\hline
&  &  \\[-5pt] 
& set & $<z>$ \\[2pt] \hline
&  &  \\[-8pt] 
$a$ & 1-4 & 3.47 \\ 
$b$ & 2-5 & 3.28 \\ 
$c$ & 4-7 & 2.99 \\ 
$d$ & 6-9 & 2.86 \\ 
$e$ & 8-11 & 2.51 \\ 
$f$ & 10-13 & 2.05 \\[2pt] \hline
\end{tabular}
\end{center}
\end{table}

\begin{figure*}
\caption[1]{The average power spectra (in units of the noise) of the Lyman-
$\alpha$ 
log$N_{HI}\geq 13.8$ (right panels) and log$N_{HI}\geq 13.3$ (left panels)
 for the
high and low redshift samples. The dotted lines are the one and two sigma
Poisson errors. The abscissa is the wavelength
$\lambda=2\pi/k$.}
 \epsfysize=12cm 
\vspace{-2cm}
\centerline{\epsffile{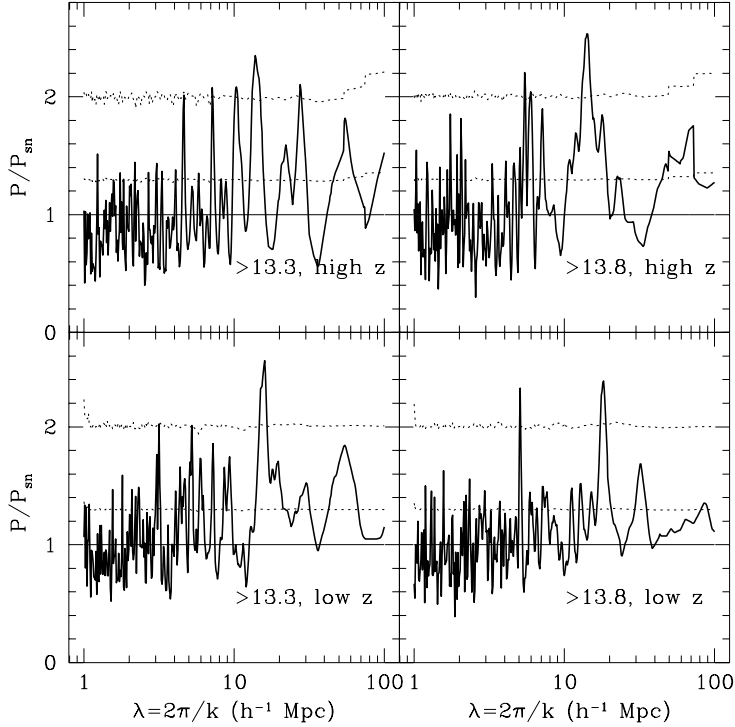}}
\label{f1}
\end{figure*}

\begin{figure*}
\caption[1]{The cumulative distribution function (circles) for all the
peaks in the spectra of our sample. The thin line is the CDF
of the Poisson simulations (an exponential), and the dashed lines mark the
 one sigma deviation.}
\label{f2} 
\centerline{\epsffile{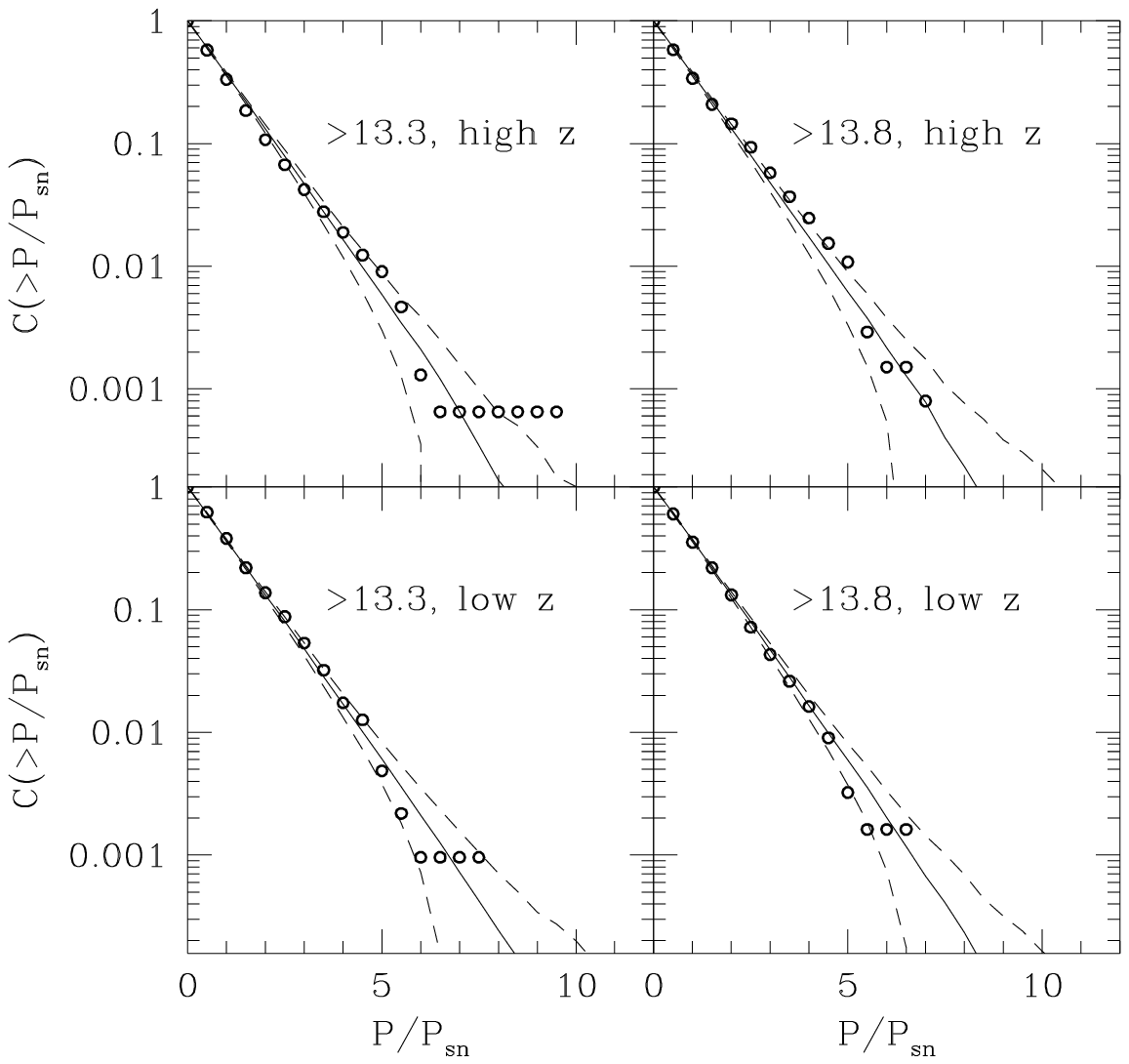}}
\end{figure*}

Let $z_i,i=1,..n$ be the redshift of the $i$-th Lyman-$\alpha $ cloud
along a line of sight, and let $\eta _i$ be its mass. We will later
assume a unitary mass for the clouds.  The Fourier transform of the
cloud distribution is then
\begin{equation}
\delta _k=L^{-1}\int [\rho (r)/\rho _0-1]\exp (ikr)dr
\end{equation}
where 
\begin{equation}
r=\frac{1}{H_0^{-1}\Omega_0^2(1+z)^2}
[2\Omega_0 z +2(\Omega_0-2)(\sqrt{1+\Omega_0 z}-1)]
\end{equation}
 is the cloud distance, $L$ is the total length of the cloud system,
$\rho (r)$ is the mass density, $\rho _0$ its average. We can put
$\rho (r)=\sum_i\eta _i\delta^D (r_i)$, where $ \delta^D (r_i)$ is a
Dirac delta function. It follows that the total mass along the line of
sight is $M=\sum \eta _i$ , and that $\rho _0=M/L=\sum \eta _i/L=\eta
_0n/L$, where $\eta_0=\sum\eta_i/n$ is the average mass. Then
\begin{equation}
\delta _k=n^{-1}\sum (\eta _i/\eta _0)\exp (ikr_i)-W_k
\end{equation}
where 
\begin{equation}
W_k=L^{-1}\int_0^L\exp (ikr)dr\simeq N_b^{-1}\sum_j^{N_b}\exp (ikr_j)
\end{equation}
is the window transform. The window integral is approximated in $N_b$
equally spaced bins, which should be taken as small as possible, down to the
resolution limit of the data. Finally, the power spectrum is defined as 
\begin{equation}
P(k)=L|\delta _k|^2
\end{equation}
By this definition, we have $P(0)=0$. If the clouds are uncorrelated, then
on average $\langle \eta _j\eta _i e^{ik(r_i-r_j)}\rangle =0$ for $i\neq j$;
therefore, we get 
\begin{equation}
\langle |\sum_j^n\eta _j\exp (ikr_j)|^2\rangle =\sum_j^n\eta _j^2
\end{equation}
and then we obtain the pure shot-noise power spectrum $P_{sn}$ in the
limit of an infinite sample (i.e. for $W_k=\delta^D(k)$)
\begin{equation}
P_{sn}=L\sum_{}^{}\eta _j^2/(\sum_{}^{}\eta _j)^2  \label{expnoi}
\end{equation}
We derived the equations including the mass, but in the following we
put $\eta _j=1$ (from which $P_{sn}=L/n$) because the relation between
the observed HI column density and the mass is dominated by poorly
known parameters like the cloud size and shape.  As a simple test, we
arbitrarily assumed $\eta _j\sim {\rm log} N_{HI}$ and found no
qualitative difference.  We will evaluate $P(k)$ at $ k_j=2\pi
j/L,j=1,..N_b/2$, so as to obtain $N_b/2$ coefficients $P(k_j)$.

The Fourier coefficients $\delta _k$, being linear combinations of many
independent random variables $\eta _i\exp (ikr_i)$, tend to be distributed
as Gaussian variables. The power spectrum coefficients, sum of the square of
two Gaussian variables, are distributed as a $\chi ^2$ distribution with two
degrees of freedom, i.e. as an exponential 
\begin{equation}
D(P_j)=(1/P_{sn})\exp (-P_j/P_{sn})  \label{pdf}
\end{equation}
If the central limit theorem conditions are not fulfilled, e.g. the
$k$-modes are correlated or the data are strongly non-Gaussian, the
distribution of $P_j$ deviates from the exponential. To estimate the
noise level $P_{sn}$ for a given line of sight, we generate one
thousand simulated distributions of lines, distributing randomly the
lines. Then, we estimate $P_{sn}$ by fitting the distribution of power
spectrum coefficients with the exponential (\ref{pdf} ), and average
over the realizations. The agreement with the expected value (
\ref{expnoi}) is excellent. From Eq. (\ref{pdf}) we see that the
variance of $P_{sn}$ is
\begin{eqnarray}
\sigma _{P_{sn}}^2&=&\int_0^{\infty} {(P-P_{sn})^2\over P_{sn}}
\exp (-P/P_{sn})dP=\nonumber\\
&&P_{sn}^2\int_0^{\infty} (x-1)^2
\exp (-x)dx=P_{sn}^2
\end{eqnarray}
However, this gives only an idea of the spread of the spectrum
coefficients around the noise. To be more quantitative, we compare the
cumulative distribution function (CDF) of the spectrum coefficients
$P_i=L|\delta _{k_i}|^2$ with the CDF of the simulated
distribution. The noise CDF is
\[
C(>P)=\int_P^{\infty}D(P^{\prime })dP^{\prime }=\exp (-P/P_{sn})
\]
Therefore, a peak of height $P=cP_{sn}$ among $n_p$ peaks is
statistically significant if $n_p\exp(-c)$ is very low,
e.g. $10^{-3}$.

\begin{figure}
\caption[1]{Theoretical trend of the conditional density for various values
of the parameters. The thick line is for a CDM model in a flat space with shape
parameter $\Gamma=0.2$. The dotted line includes the non-linear correction.
The thin line assumes a higher value of the velocity dispersion. 
The short-dashed
line is for $\Gamma=0.5$, while the long-dashed one 
is for a open model. All models
are normalized to $\sigma_8=0.6\Omega_0^{-0.6}$, and are evaluated at
$z=3$. Beyond 3 Mpc/$h$ the non-linear correction becomes negligible.}
\epsfysize=6.5cm
\centerline{\epsffile{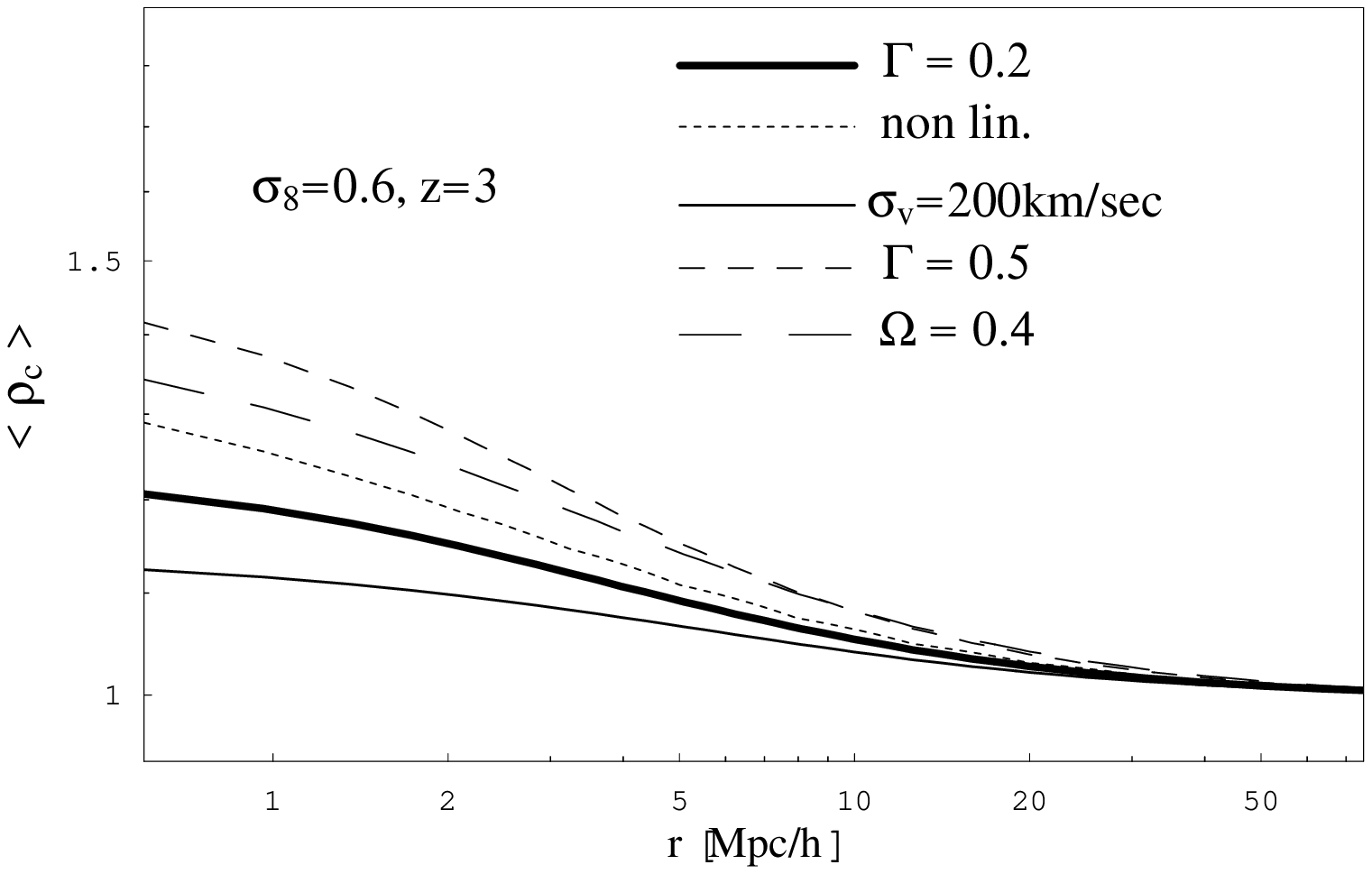}}
\label{f3}
\end{figure}

\begin{figure}
\caption[1]{The circles represent the normalized integral density of
neighbours for the seven high redshift cloud systems with various cut-offs in
column density. The short-dashed and the long-dashed lines give the
one-sigma and the two-sigma error band, respectively. The continuous lines
are the linear CDM expectations, for $\Gamma =0.2$, the lower curve, and $
\Gamma =0.5,$ the upper curve. The upper and the lower panels are for the
seven high and low redshifts respectively.}
\label{f4}
\epsfysize=8.9cm 
\centerline{\epsffile{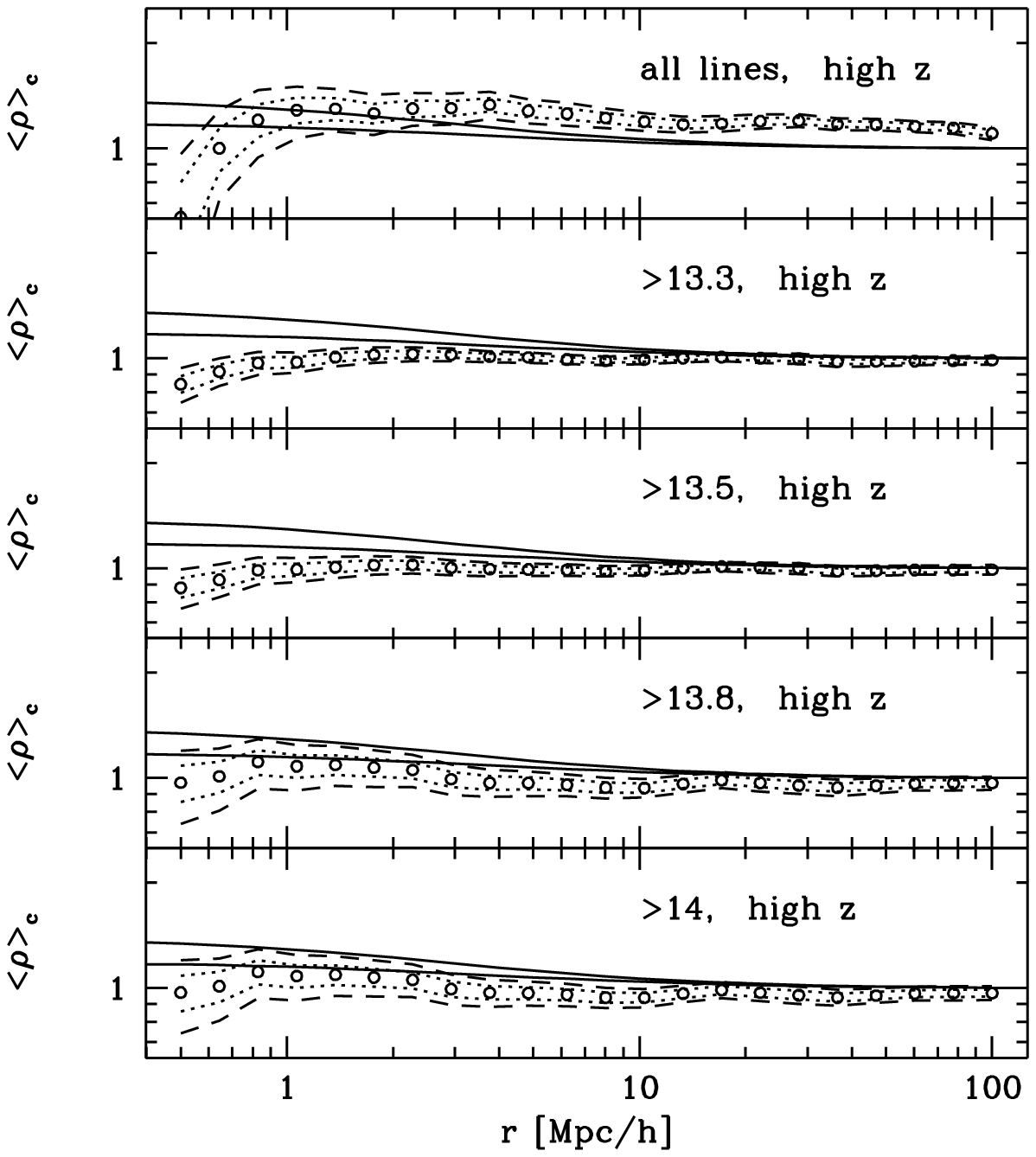}} 
\epsfysize=8.9cm 
\centerline{\epsffile{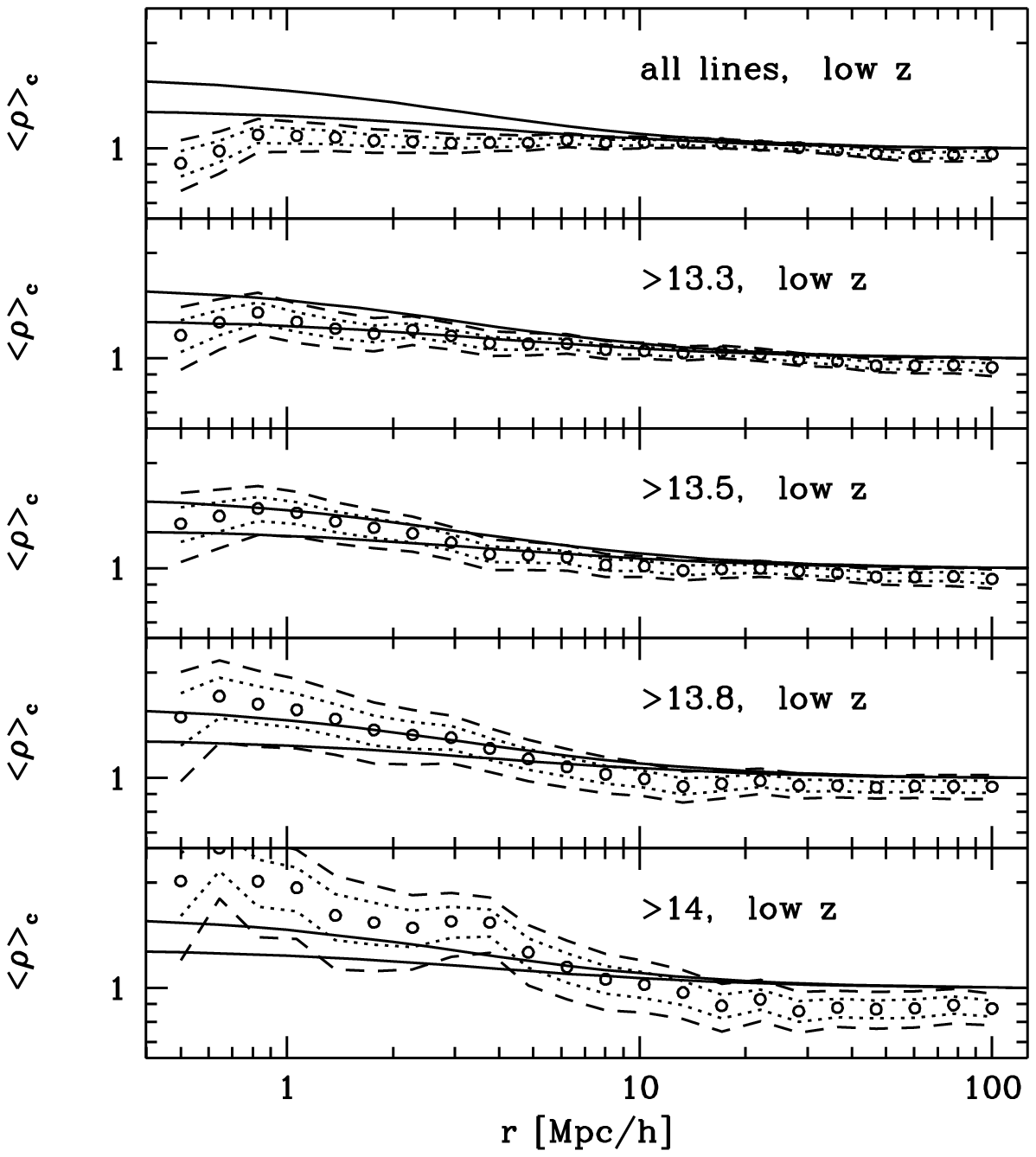}}
\end{figure}

\begin{figure}
\caption[1]{As in Fig. 4b, now assuming an open universe
with
$\Omega_0=0.4$.}
\label{fig4omega}
\epsfysize=8.9cm 
\centerline{\epsffile{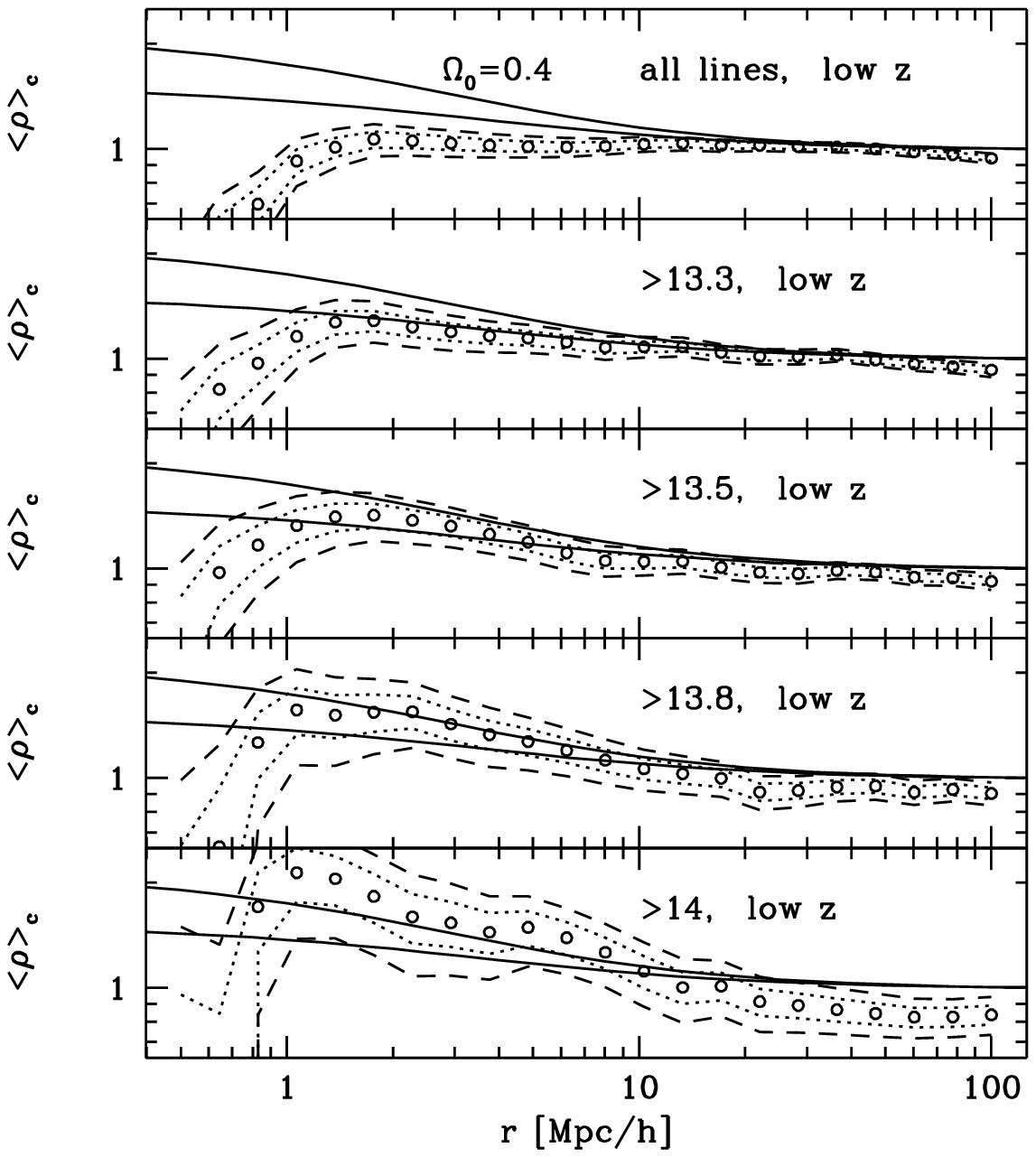}}
\end{figure}

The one-dimensional power spectrum is plagued by the aliasing problem
stressed by Kaiser \& Peacock (1991): spurious high signal-to-noise
peaks may be induced by the 1-dim geometry.  However, since we compare
the Ly$\alpha$ spectrum to random realizations with the same geometry,
and not to a theoretical significance level, our method is able
to assess the probability that random realizations produce a given
spectrum peak.

The second quantity we use is the normalized density of neighbours, i.e. 
\begin{equation}
\rho _n=\langle \rho (R)\rangle _n=\langle \sum_i\eta _i/R\rangle /\rho _p
\end{equation}
where the sum extends to all the lines within a distance $R$ of a
given line assumed as a center, and the average extends over all the
centers, and where $\rho_P=n/L$ is the expected Poisson density. At
the scales at which a clustering signal is detected, $\rho _n>1.$

Evaluating $\rho _n$ the following problem arises. Let us order the
Lyman-$\alpha$ clouds along a line of sight on the $r$ axis, with
increasing distance $r$.  Consider the Lyman-$\alpha$ clouds within a
distance $R$ from the $i$-th cloud at position $r_i$: there are clouds
on the right, at $r_i < r \leq r_i+R$ , and clouds on the left, at
$r_i-R \leq r < r_i$. Suppose now the left set, for instance, is cut
by the boundary of the sample.  We would count the clouds contained in
it in $\rho(R)$ even if, in fact, the left set contains lines only up
to a distance smaller than $R$. If the clustering decreases with
scale, as expected, the left set cut by the boundary would have on
average a higher $\rho(R)$ than if it were not cut. In other words,
including sets cut by the boundary amounts to averaging different
scales in $\rho(R)$.  This would result in a systematic boundary
effect, an effect which increases for large radii and for large
clustering.  To avoid this, we include only the clouds in sets which
are not cut by the boundary: that is, around each cloud, we take only
the right set if the left one is cut, and vice versa.

Let us write down here the relations between the two quantities introduced,
the power spectrum and the density of neighbours, and the relation between
these and the 3-dim power spectrum $P_3(k)$. We have 
\begin{equation}
P(k)=\frac 1{2\pi }\int_k^\infty P_3(k^{\prime })k^{\prime }dk^{\prime }
\end{equation}
\begin{equation}
\rho _n=1+\frac 1{2\pi }\int_0^\infty P(k)W(kR)dk  \label{p1rho}
\end{equation}
where $W(x)$ is the 1-dim top-hat window function, $W(x)=\sin (x)/x$.

The theoretical spectra are in real space. As is well known, the
peculiar velocities suppress the power of the real space spectrum at
small scales and enhance it at large scales. The redshift space
spectrum $P_{3s}(k)$ is then a function of the real space spectrum
$P_{3r}(k),$ of the factor $\beta =\Omega_0 ^{0.6}$ , and of the cloud
velocity dispersion $\sigma_v$ along the line of sight.  The overall
effect can be modeled according to the following formula derived by
Peacock \& Dodds (1994):
\begin{eqnarray}
P_{3s}(k) &=&P_{3r}(k)G(\beta ,y )  \nonumber \\
G(\beta ,y ) &=&\frac{\pi ^{1/2}}8\frac{{\rm erf}(y )}{y
^5}[3\beta ^2+4\beta y^2+4y^4]-  \nonumber \\
&&\frac{\exp (-y^2)}{4y^4}[\beta ^2(3+2y
^2)+4\beta y^2]  \label{dist}
\end{eqnarray}
where $y=k\sigma_v H_0^{-1}$.  We will investigate scales from 1 to
100 Mpc/$h.$ The lower limit of the scale is close to the typical
transverse size of the {\rm Ly}$\alpha $~clouds, being of the order of
a few hundred $h^{-1}$ kpc with no strong redshift evolution (Crotts
et al. 1997, Smette et al. 1995).  At scales below 1 or 2 Mpc/$h$ it
is very difficult to model accurately the Lyman-$\alpha$ power
spectrum, due to various effects: the line-of-sight velocity
dispersion, the thermal broadening of the lines, the procedure of line
fitting, the non-linear clustering. Croft et al. (1998) found that a
correction factor $\exp (-.5 k^2 r_s^2)$ with $r_s$=1.5 Mpc/$h$ may
account empirically for all these small scale effects. Although we
included some of the effects above mentioned to correct the spectrum
at small scales, we work out the numerical results using only scales
larger than 3 Mpc/$h$, where the corrections are small.  We take for
the velocity dispersion $\sigma_v =100 $ \kms: simulations give in
fact peculiar velocities about 100 km~s$^{-1}$ or less
(Miralda--Escud\`{e}, 1996, Rauch et al. 1997, Bi \& Davidsen 1997),
while observations of quasar pairs give velocity differences between
common absorption lines less than about 100 km~s$^{-1}$~(Smette et
al., 1995, Dinshaw et al., 1994, $z\sim 2$).  Although the level of
peculiar velocity is expected to vary with redshift, we fix its value
to 100 km/sec both because the range of $z$ we investigate is not very
large, and because the correction is anyway important only at scales
less than 1 Mpc/$h$.

Let us mention that the expected IGM spectrum deviates
from the CDM spectrum at small scales also because of the clustering
suppression below the IGM Jeans mass. This deviation can be accounted
for by a term which depends essentially on the IGM temperature and
redshift (Bi \& Davidsen 1997).  As long as $T<10^5$K (Kim et al. (1997)
found that at $z=3$ less than 19\% of the \lya~clouds with $12.8
\leq \log N_{HI} \leq 16$ can have $T>10^5$K), the scale at which a
sensible deviation occurs is smaller than 0.5 Mpc $h^{-1}$, for all
relevant $z$. In the following we therefore neglect this factor.

\section{Power spectrum results}

 We calculated the PS for all our cloud systems, and averaged over the
high and low redshift subsets mentioned in Section 2. In this section
we assume for simplicity $\Omega_0=1$, since here we will not compare
the results to a CDM model, but rather to Poisson distributions.  We
have tested that putting $\Omega_0=0.4$ in the distance-redshift
relation does not modify the conclusions.  In Fig.~\ref{f1} we show
the power-to-noise ratios $P(k)/P_{sn}$ for the average of the six
higher and, separately, of the seven lower redshift QSO lines of sight
in the range from 1 to 100 Mpc $h^{-1}$with cut-offs $\log
N_{HI}>$13.3 and $\log N_{HI}>$13.8 (the wavelength being $2\pi
/k$). Since we have expressed our spectra in units of the noise, we
can take the average straightforwardly: the difference in line density
due to evolution and to possible other observational effects is
automatically taken into account. The result is that there is no
evidence of deviation from the Poisson spectra, i.e. that there are no
preferential scales, or ''periodicity'', in the clustering. Although
there are a few isolated peaks above 2$\sigma $, they are not stronger
and more frequent than one would get in a pure Poisson
distribution. This is quantitatively reported in Fig.~\ref{f2}, where
we display the cumulative distribution of the peaks in all the 13
spectra at two column density cut-off: the CDF is always less than
1$\sigma $ from Poisson, even if we can have signals as strong as 9.5
times the noise (top left panel of Fig.~\ref{f2}).  The same holds
true for all the other cuts in column density.

We conclude that the power spectrum analysis of the distribution of
\lya~lines does not detect clustering on any particular scale. This
does not necessarily mean, of course, that there is no clustering
above Poisson; a suitable filtering of the data could in fact reveal a
significant signal. However, a filtering in $k$-space has not in
general a direct geometrical meaning. It is therefore simpler to
consider an integral statistics in real space, such as the integral
density of neighbours, and to compare this statistics with the
theoretical expectation.

\section{Density of neighbours results.}

As one can see from the results of the previous Section, the 1-dim power
spectrum is extremely noisy, and cannot easily be compared with the
theoretical predictions. To perform this comparison, we adopt then in this
Section the integral density of neighbours.

The theoretical linear CDM-like galaxy spectrum in real space can be
written as
\[
P_{3r,cdm}\left( k,\Gamma \right) =AkT(k,\Gamma )^2
\]
where $T(k,\Gamma )$ is the transfer function of Davis et al. (1985),
$\Gamma $ is a theoretical free parameter (equal to $\Omega_0h$ in CDM
models if $k$ is in $h$ Mpc$^{-1}$) and $A$ is the normalization
factor.  We normalize putting the present variance in 8 Mpc $h^{-1}$
spherical top-hat cells equal to the cluster abundance value
$0.6\Omega_0^{-0.6}$ (White, Efstathiou \& Frenk 1993).  We corrected
the spectrum for the non-linear enhancement as suggested by Peacock \&
Dodds (1996), but we found that at scales larger than 3 Mpc/$h$ the
effect is negligible. As anticipated, we will compare quantitatively
our results to the data only at scales larger than 3 Mpc/$h$.  We have
to determine the expected one-dimensional spectrum from the 3D
spectrum in redshift space, $P_{3s,cdm}\left( k\right) ,$ and to
evolve it back. This gives the spectrum
\begin{equation}\label{back}
P_{cdm}(k,z)=D_+(\Omega_0,z)^2\int_k^{\infty}
k^{\prime }dk^{\prime }P_{3s,cdm}\left(
k^{\prime } \right) 
\end{equation}
Putting $P_{cdm}$ in Eq. (\ref{p1rho}) we can obtain the expected
density of neighbours. In Fig.~\ref{f3} we report the theoretical
density of neighbours for some interesting cases.  Naturally, the
expression (\ref{back}) is an oversimplification. Both the non-linear
and the redshift corrections induce $z$-dependent distortions on the
spectrum. An accurate modeling of these distortions requires extensive
$N$-body calculations.  However, these distortions are important only
at small scales, where in any case the uncertainties in the estimation
of the Lyman-$\alpha$ are a major hurdle to a precise comparison.

\begin{figure}
\caption[1]{The density of neighbours for the samples $a-f$ listed in Table 
\ref{t2}, and for $\log N_{HI}\geq 13.8$.}
\epsfysize=9cm 
\centerline{\epsffile{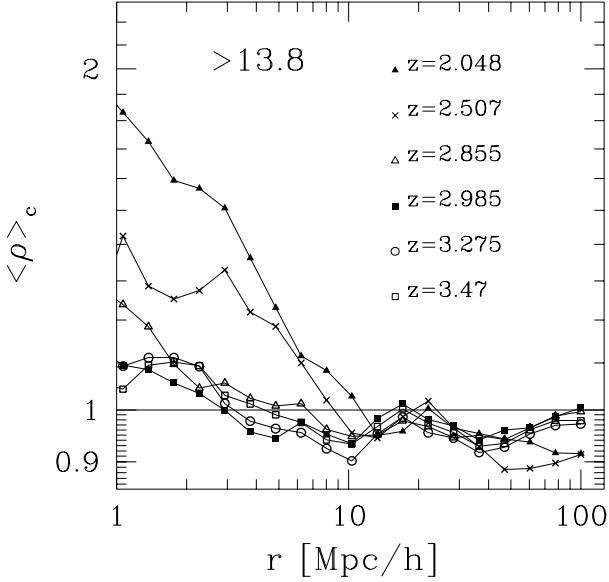}}
\label{f5}
\end{figure}

According to the standard model of biased gravitational instability,
and to the findings of Cristiani et al. (1996), we expect that the
clustering increases with decreasing redshift, and with increasing
column density.  However to fully understand how the 1D distribution
of HI column density in QSO spectra represents the true 3D
distribution of hydrogen in the Universe at different redshifts is not
an easy task. Given an HI column density associated with an absorption
line, the real mass of the {\rm Ly}$\alpha $ ~clouds depends on the
ionisation correction $N_{HI}/N_H$ in the cloud which gives the total
hydrogen column density, and on the geometry of the cloud.  Both these
quantities can be function of redshift due to the evolution of the UV
background flux $J_\nu,$ which photoionises the cloud and to the
evolution of the gravitational instabilities in the clouds. To
estimate a pure redshift evolution of the 3D power spectrum of the
intergalactic HI, it is thus necessary to consider any redshift
evolution of the {\rm Ly}$\alpha$ clouds not associated with
gravitational clustering. Since this is not a trivial problem, in this
paper we do not investigate the evolution of the clustering in
redshift at a fixed value of the column density, but, as a preliminary
task, we map the variation of clustering both with redshift and column
density. We leave to future work the comparison of this evolution map
with the theory or with $N$-body and hydrodynamic simulations, a
necessary step in the reconstruction of the history of the IGM.

We show in Figs.~\ref{f4} the density of neighbours for several of our
subsets. The error bands are Poisson errors estimated via MonteCarlo
simulations; we adopted the Poisson errors since the level of cosmic
variance is always negligible.  For the set of the six highest
redshift systems ($<z>=3.28$) there is no detectable clustering, no
matter what the cut in the column density is. For the lowest redshift
subset ($<z>=2.39$), we see a clear increase in clustering with $
N_{HI}$. In Fig. (\ref{fig4omega}) the same low redshift subset is
shown, now assuming an open universe with $\Omega_0=0.4$.  Comparing
with the CDM curve, we see that the lines with $\log N_{HI}\geq 13.3$
are near or above the CDM linear predictions, both for the flat and
the open geometry. We also compared the results for the two line sets
1 and 1$^b$ listed in Table \ref {t1}; the results are compatible
within the errors.  It is interesting to observe that the CDM model is
not an adequate fit for the low $z$, high column density data,
independently of the bias factor. The normalized density of neighbours
is in fact always significantly less than unity at large scales (that
is, there is large scale anticorrelation) contrary to the CDM model.
We could not reproduce accurately this large scale feature even by
varying the shape parameter $\Gamma$.

\begin{figure}
\caption[1]{The bias function $b_{{\rm {Ly}\alpha }}^2(z, N_{HI})$,
assuming $\Omega_0=1$.}
\label{f6}
\epsfxsize=9cm 
\epsfysize=9cm 
\centerline{\epsffile{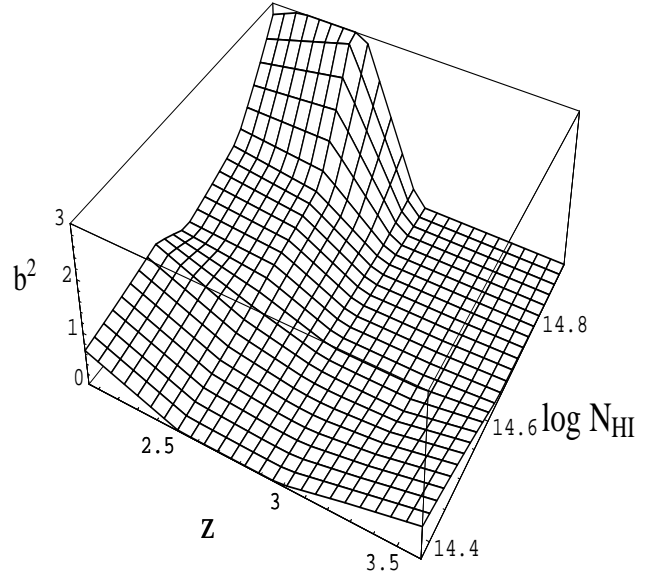}}
\end{figure}

In Fig.~\ref{f5} we show the clustering trend in the six subsets of
Table \ref{t2}, for $\log N_{HI}\geq 13.8$, assuming $\Omega_0=1$.
The clustering increases regularly from $z\simeq 3.5$ to $z\simeq 2 $,
although the increase is considerably faster at low $z$.  Assuming as
a reference the CDM predictions with $\Gamma =0.2$ and linear
evolution, we can estimate from the sets in Table \ref{t2} the biasing
function
\begin{eqnarray}
b_{{\rm Ly}\alpha }^2(z, N_{HI}) &=&{\rho _n(z, <N_{HI}>)-1\over
\rho
_{n,CDM}(z)-1}=  \nonumber \\
&&P_{{\rm Ly}\alpha }(k)/P_{cdm}(k)
\end{eqnarray}
which expresses for any given average $ <N_{HI}>$ and any given
redshift, the relation between the real clustering and the linear CDM
predicted clustering (the average $ <N_{HI}>$ has been obtained
excluding the very few lines with $\log H_{HI}>16$). Clearly, $b_{{\rm
{Ly}\alpha }}\geq 1$ means that the data are more clustered than the
linear-extrapolated CDM distribution; $b_{Ly\alpha }\leq 1$ implies an
under-clustering of the data. In Fig.~\ref{f6} we plot the function
$b_{{\rm {Ly}\alpha }}^2(z, N_{HI})$.  As already mentioned, the real
evolution is likely to be both in column density and clustering
strength. If one assumes arbitrarily that the average column density
does not change with time, then the evolution in redshift turns out to
be extremely fast. In flat space, clustering growth as fast as
$(1+z)^{-3}$ or faster seem to fit the trend we obtain for high column
density cut-offs.  One can obtain more reasonable values, i.e. values
of closer to the linear trend {\it only} by assuming at the same time
a decrease of the average column density with decreasing redshift.  To
give a rough approximation, mostly valid at redshift near 2, we can
fit the function $b_{{\rm {Ly}\alpha }}$ as
\begin{equation}
b_{{\rm {Ly}\alpha }}=
\left(\frac{ N_{HI}}{N_1}\right)^{0.65}\left(\frac{1+z}{1+z_1}\right)^{-1.7}  
\label{fit}
\end{equation}
where $z_1$=3, and $N_1=7.1\cdot 10^{14}$ atoms/cm$^2$
if $\Omega_0=1$, and as
\begin{equation}
b_{{\rm {Ly}\alpha }}=
\left(\frac{ N_{HI}}{N_1}\right)^{0.6}\left(\frac{1+z}{1+z_1}\right)^{-1.4}  
\label{fit04}
\end{equation}
where $z_1$ is the same, and $N_1=9.7\cdot 10^{14}$ atoms/cm$^2$ for
$\Omega_0=0.4$. The relative error in the exponents is in all cases of
the order of 30\%.  The bias function shows quantitatively the
clustering trend both in column density and redshift.  As an
interesting consequence, we see that the clustering evolves linearly,
i.e. $b_{{\rm {Ly}\alpha }}$ is constant in redshift, only if $
N_{HI}\sim (1+z)^{1.7/0.65}\sim (1+z)^{2.6\pm 1}.$ (or as
$(1+z)^{2.3\pm 1}$ for the open case).  Notice that this conclusion is
independent of the overall amplitude of the bias function; it depends
only on the relative level of clustering of the Ly$\alpha$
distributions.

Two caveats have to be kept in mind.  First, our assumption of a
scale-independent bias may prove incorrect. However, we used it as a
simple way to parameterize the Lyman-$\alpha$ spectrum in terms of the
CDM spectrum, and as such it looks, {\it a posteriori}, as an
acceptable approximation.  Second, the clustering evolution of the
clouds need not be linear, even at the scales we are investigating. In
fact, along with the linear evolution due to the gravitational
instability, the clustering may evolve also in response to the
evolution of the ionising sources.  If, as shown e.g. by Bagla (1998),
the evolution of the clustering of the first collapsed objects is far
from linear at $z>1$, then the Lyman-$\alpha$ clouds evolution could
also be non-linear.

\section{Conclusions}

The aim of this paper was to reconstruct the power spectrum of the
Lyman-$\alpha $ clouds as a function both of redshift and average
column density, and to compare it to theoretical CDM models.  We first
estimated directly the line-of-sight power spectrum: we concluded that
there is no evidence of characteristic scales in the cloud
distribution, contrary to what is found in some galaxy surveys.  Since
the direct reconstruction of the power spectrum is too noisy to derive
useful comparisons with theoretical expectations, we employed next the
integrated density of neighbours. We assumed that the cloud spectrum
can be fitted by
\begin{equation}
P_{{\rm Ly}\alpha }(z,k)=b_{{\rm Ly}\alpha }^2(z, N_{HI})P_{cdm}(z,k)
\label{parform}
\end{equation}
where $b_{{\rm Ly}\alpha }^2$ characterizes the Lyman-$\alpha$ line
bias with respect to a CDM model normalized to the cluster abundance.
This fit performs acceptably well for scales between 1 and 10 Mpc$/h$,
while the data are systematically anticorrelated on larger scales.  We
found that the Ly$\alpha $ clouds are clustered with power similar to
the theoretical CDM only at redshift near 2 and average column density
higher than 10$^{13.3}$ HI atoms/cm$^2$, being underclustered
($b_{{\rm Ly}\alpha }^2<1$) in all the other cases. A reasonable
expectation is that the cloud evolution is linear for most of the
scales considered here. Imposing this condition upon the fits
(\ref{fit}) and (\ref{fit04}) we conclude that the average column
density must decrease in time as $ N_{HI}\sim (1+z)^{2.6}$ (or as
$(1+z)^{2.3}$ for a $\Omega_0=0.4$ universe).  For the popular model
with a cosmological constant such that $\Omega_{\Lambda}=0.6$ the
results of the open case applies to a good approximation, since
$\Omega_{\Lambda}$ does not affect significantly the clustering
evolution. Notice that the function $b_{{\rm Ly}\alpha}$ depends on
assuming the specific parametric form (\ref{parform}) and in
particular on the clustering redshift evolution of the dark matter
spectrum. However, the trend in $z$ we find is so fast that it seems
difficult to account for it without a strong redshift decrease of
$N_{HI}$.

It is clear that the dataset investigated here is too small to give
definitive results. One of the problems is that the most recent and higher
cut-off samples are also the most sparse, and thus give higher statistical
uncertainty. One conclusion is however rather firm:   the cloud clustering can
be explained only by assuming a simultaneous increase in correlation and
decrease in average column density with redshift.

\end{document}